\documentclass[a4paper,twoside]{article}

\usepackage{epsfig}
\usepackage{subcaption}
\usepackage{calc}
\usepackage{amssymb}
\usepackage{amstext}
\usepackage{amsmath}
\usepackage{amsthm}
\usepackage{multicol}
\usepackage{pslatex}
\usepackage[round]{natbib}
\usepackage{algorithm2e}
\usepackage[bottom,hang,flushmargin]{footmisc}
\usepackage{SCITEPRESS}

\usepackage{booktabs}
\usepackage{array}
\usepackage[table]{xcolor} %
\usepackage{multirow}       %
\usepackage{graphicx}       %
\usepackage{makecell}
\usepackage{adjustbox}
\usepackage{float}
\usepackage{pifont}

\usepackage[most]{tcolorbox}
\usepackage{enumitem,kantlipsum}
\newcommand{\customlabel}[1]{\small{\textbf{C#1.}}}
\newtcolorbox{contributionsbox}{title=Contributions, boxrule=1.25pt, left=2pt, top=1pt, bottom=0.5pt}

\newenvironment{contributions}{%
  \mbox \\%
  \begin{contributionsbox}%
    \begin{enumerate}[leftmargin=0.55cm, label=\customlabel{\arabic*}]%
      }{%
    \end{enumerate}%
  \end{contributionsbox}%
}

\usepackage{hyperref}

\definecolor{yescolor}{RGB}{102, 205, 170} %
\definecolor{nocolor}{RGB}{255, 165, 79}   %
\newcolumntype{M}[1]{>{\centering\arraybackslash}m{#1}}

\usepackage{glossaries}
\usepackage{glossaries-extra}
\setabbreviationstyle[acronym]{long-short}
\glsdisablehyper

\newacronym{oc}{OC}{organic computing}
\newacronym{dl}{DL}{deep learning}
\newacronym{ssl}{SSL}{self-supervised learning}
\newacronym{sl}{SL}{supervised learning}
\newacronym{ml}{ML}{machine learning}
\newacronym{tl}{TL}{transfer learning}
\newacronym{ws}{WS}{weak supervision}
\newacronym{al}{AL}{active learning}
\newacronym{dal}{DAL}{deep active learning}
\newacronym{mlp}{MLP}{multi-layer perceptron}
\newacronym{dnn}{DNN}{deep neural network}
\newacronym{sota}{SotA}{state-of-the-art}
\newacronym{cv}{CV}{computer vision}
\newacronym{plm}{PLMs}{pretrained language models}
\newacronym{cnn}{CNN}{convolutional neural network}
\newacronym{rnn}{RNN}{recurrent neural network}
\newacronym{ast}{AST}{audio spectrogram transformer}
\newacronym{w2v2}{W2V2}{Wav2Vec 2.0}
\newacronym{fsl}{FSL}{Few-Shot Learning}
\newacronym{rl}{RL}{Reinforcement Learning}
\newacronym{hf}{HF}{hugging face}
\newacronym{vit}{ViT}{vision transformer}
\newacronym{lora}{LoRA}{Low-Rank Adaptation}
\newacronym{llm}{LLM}{large language model}
\newacronym{mc}{MC}{Monte Carlo}

\newacronym{pam}{PAM}{passive acoustic monitoring}
\newacronym{sed}{SED}{sound event detection}
\newacronym{bsc}{BSC}{bird sound classification}
\newacronym{snr}{SNR}{signal-to-noise ratio}
\newacronym{stft}{STFT}{Short-Time Fourier Transform}

\newacronym{ts}{TS}{temperature scaling}
\newacronym{ps}{PS}{Platt scaling}
\newacronym{nll}{NLL}{negative log-likelihood}

\begin{document}

\title{Uncertainty Calibration of Multi-Label Bird Sound Classifiers}

\author{
  \authorname{
    Raphael Schwinger\sup{1}\orcidAuthor{0009-0001-8519-3571}, Ben McEwen\sup{2}\orcidAuthor{0000-0002-0869-7717},
    Vincent S. Kather \sup{2,3}\orcidAuthor{0000-0003-1491-0964},
    René Heinrich \sup{4,5}\orcidAuthor{0000-0002-1939-501X},\\
    Lukas Rauch \sup{5}\orcidAuthor{0000-0002-6552-3270},
  and Sven Tomforde\sup{1}\orcidAuthor{0000-0002-5825-8915}}
  \affiliation{\sup{1} INS, Kiel University, Germany}
  \affiliation{\sup{2} Tilburg University, Netherlands}
  \affiliation{\sup{3} Naturalis Biodiversity Center, Netherlands}
  \affiliation{\sup{4} Fraunhofer IEE, Germany}
  \affiliation{\sup{5} IES, University of Kassel, Germany}
  \email{raphael.schwinger@cs.uni-kiel.de}
}

\abstract{
  Passive acoustic monitoring enables large-scale biodiversity assessment, but reliable classification of bioacoustic sounds requires not only high accuracy but also well-calibrated uncertainty estimates to ground decision-making. In bioacoustics, calibration is challenged by overlapping vocalisations, long-tailed species distributions, and distribution shifts between training and deployment data. To the best of our knowledge, the calibration of multi-label deep learning classifiers within the domain of bioacoustics has not yet been systematically assessed. We systematically benchmark the calibration of four state-of-the-art multi-label bird sound classifiers on the BirdSet benchmark, evaluating global, per-dataset, and per-class calibration using threshold-free calibration metrics (ECE, MCS) alongside discrimination metrics (cmAP).
  Model calibration varies significantly across datasets and classes. While Perch v2 and ConvNeXt$_{BS}$ show better global calibration, results vary between datasets. Both models indicate consistent underconfidence, while AudioProtoPNet and BirdMAE are mostly overconfident.
  Surprisingly, calibration seems to be better for less frequent classes. Using simple post hoc calibration methods we demonstrate a straightforward way to improve calibration. A small labelled calibration set is sufficient to significantly improve calibration with Platt scaling, while global calibration parameters suffer from dataset variability. Our findings highlight the importance of evaluating and improving uncertainty calibration in bioacoustic classifiers.
}
\onecolumn \maketitle \normalsize \setcounter{footnote}{0} \vfill

\begin{figure}[ht!]
  \centering
  \includegraphics[width=0.95\columnwidth]{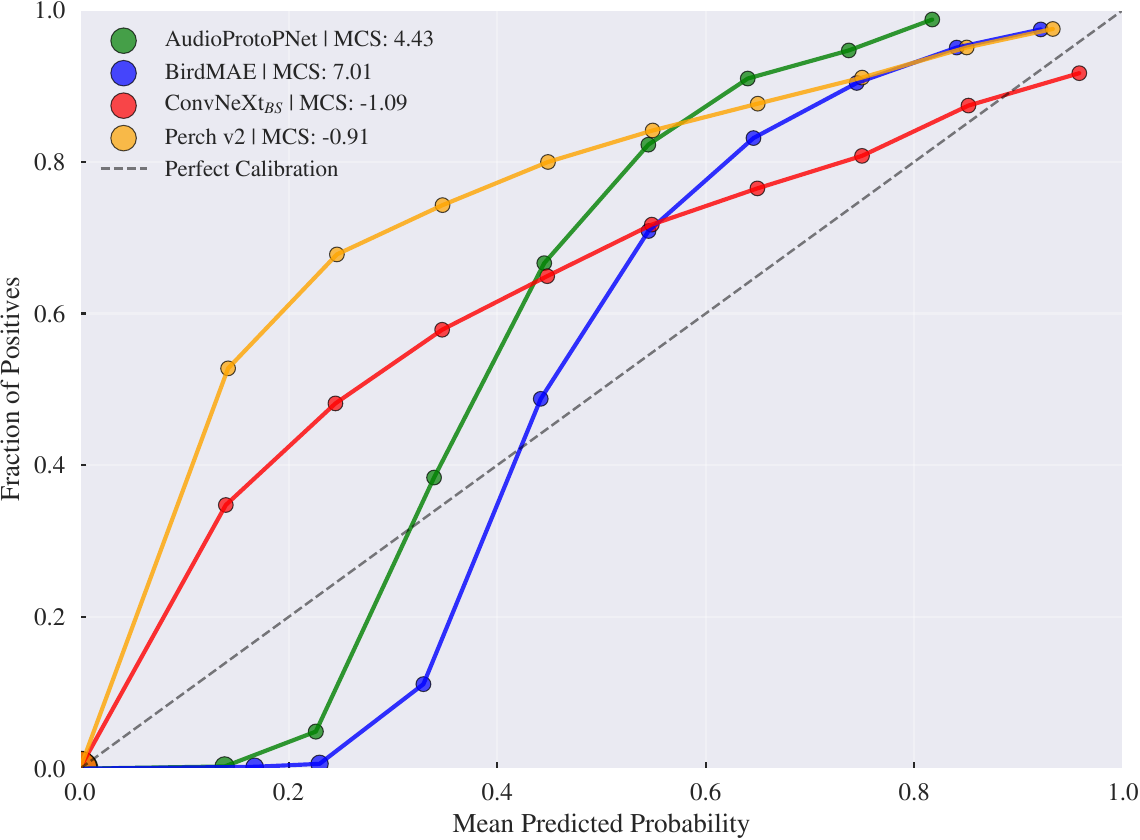}
  \caption{Calibration evaluation of AudioProtoPNet, BirdMAE, ConvNeXt$_{BS}$ and Perch v2  globally on all BirdSet test samples visualised in a reliability diagram. Mean predicted probabilities are shown on the x-axis, while empirical frequencies are shown on the y-axis. The miscalibration score (MCS) is outlined for each model.}
  \label{fig:reliability-diagram-combined}
\end{figure}

\section{\uppercase{Introduction}}
\label{sec:introduction}

Human pressures are leading to declines in biodiversity, necessitating the development of measurement systems and effective analysis tools to monitor these trends~\citep{schmeller_building_2017,gosselin_relationships_2018}.
Bioacoustics describes the study of sounds produced by organisms, offering a non-invasive measurement method of ecological habitats and their species~\citep{laiolo_emerging_2010}.
\Gls{pam} has emerged as a powerful technique for continuous and non-invasive biodiversity assessments of animal populations in remote areas through low-cost recorders at scale~\citep{sugai_terrestrial_2019}.
Recent advances in \gls*{dl} have made it feasible to process large collections of PAM recordings and automatically detect species vocalisations, enabling applications such as population trend estimation~\citep{mcginn_feature_2023}, impact assessments~\citep{kershenbaum_automatic_2025}, and rapid alerts for protected or invasive species~\citep{baumgartner_persistent_2019}.
While the performance of \gls*{dl} models in bioacoustics has shown steady increases in previous years, generalisation to soundscape recordings with overlapping vocalisations, as collected in \gls{pam} scenarios, is still challenging~\citep{stowell_computational_2022,rauchBirdSetLargeScaleDataset2025,schwinger2025foundation}.
In practice, decisions based on classifier outputs rarely stand alone: ecologists, regulators, and land managers need to understand when to rely on model outputs, and when to defer to human review or collect more data. Variability in model calibration across domains calls into question the validity of fixed thresholds when applied across different deployment settings~\citep{navine2024all}. Reliable decision-making therefore requires not only high discrimination (e.g., AUROC, cmAP scores) but also well-calibrated predictive probabilities that honestly reflect uncertainty~\citep{guo2017calibration}.

Calibration is especially challenging in bird sound classification as the task
is inherently multi-label due to overlapping vocalisations. In addition, soundscapes vary widely across habitats and devices, and species occurrences follow a long tail, with rare species being of greatest conservation concern~\citep{stowell_computational_2022}. Distribution shift between training data (often directional focal recordings) and deployment data (omnidirectionally recorded soundscapes) can further degrade reliability~\citep{rauchBirdSetLargeScaleDataset2025}. In these settings, overconfident models can bias occupancy predictions, inflate apparent detection rates, and mislead thresholding and decision making, leading to samples not flagged for human evaluation. Conversely, underconfidence reduces the utility of scores for prioritising expert review. Reliable uncertainty estimates are therefore critical to support transparent and accountable use of PAM in environmental decision-making.

Despite growing interest, the calibration behaviour of multi-label audio classifiers remains underexplored. Many studies prioritise accuracy or ranking metrics, while probability quality is reported sparsely or via single-number summaries that may obscure class-specific effects~\citep{merrienboerPerch20Bittern2025,kahlBirdNETDeepLearning2021}. Moreover, commonly used post hoc calibration methods, which improve model calibration without retraining, are not uniformly evaluated in this domain. Based on this motivation, our contributions can be summarised as follows:

\begin{contributions}
\item Benchmark calibration of four state-of-the-art multi-label bird sound classifiers on BirdSet, reporting global, per-dataset and per-class calibration with threshold-free metrics, revealing that per-domain (e.g., dataset, class, etc.) calibration is highly variable.
\item Investigate post hoc strategies (temperature and Platt scaling) showing that Platt scaling with 10 minutes of test data can significantly improve calibration.
\item Summarise implications to give practitioners concrete suggestions for deployment-specific calibration evaluation and improvement. Code is available online\footnote{\url{https://github.com/DBD-research-group/UncertainBird}}.
\end{contributions}

\section{\uppercase{Background}}
\label{sec:background}

\subsection{Uncertainty and Calibration}
\label{sec:uncertainty_and_calibration}
\paragraph{Definitions.} We distinguish \emph{predictive uncertainty} from \emph{calibration}. Uncertainty quantifies how unsure a model is about its prediction for an input $x$. In classification, this is reflected by the confidence score or in the spread or sharpness (entropy) of the predictive probability distribution $\hat{\mathbf p}(x)$. Calibration measures whether these stated probabilities correspond to empirical frequencies. A perfectly calibrated predictor satisfies, for each class $c$, $\Pr\big(Y_c=1\mid \hat p_c = p\big)=p$ for all $p\in[0,1]$.

\paragraph{Types of uncertainty.} We follow the common decomposition into (i) \emph{aleatoric} uncertainty inherent in the data (e.g., noise, wind, overlapping vocalisations, distant calls) and (ii) \emph{epistemic} uncertainty due to a lack of model knowledge (e.g., rare species, novel soundscapes, or data-scarce regions)~\citep{rewickiEstimatingUncertaintyDeep2022a}. While we are primarily motivated by epistemic uncertainty, the calibration methods we employ are general-purpose post hoc techniques; we assume that acquiring more data is not possible.

\paragraph{Reliability diagram.} A reliability diagram plots the predicted probabilities against the observed frequencies of the positive class. For a well-calibrated model, points should lie on the diagonal. The model is overconfident in a probability region if the points are positioned under the diagonal and underconfident if they are over the diagonal.

\paragraph{Calibration methods.}

Calibration methods can be broadly divided into post hoc and ad hoc approaches. \textit{Post hoc} methods operate on a trained model's outputs, learning a mapping from uncalibrated logits to calibrated probabilities without modifying the model weights. \Gls{ts}~\citep{guo2017calibration}, \gls{ps}~\citep{platt1999probabilistic}, isotonic regression~\citep{niculescu2005predicting}, and vector scaling~\citep{kull2019beyond} are widely used post hoc techniques. These methods are attractive due to their simplicity and low computational overhead. Bayesian deep learning methods, such as Monte Carlo dropout~\citep{gal2016dropout}, deep ensembles~\citep{lakshminarayanan2017simple}, and variational inference~\citep{blundell2015weight,rizos2024propagating}, provide principled uncertainty estimates but add computational overhead.

\textit{Ad hoc} calibration, in contrast, integrates calibration objectives directly into model training. This includes approaches such as label smoothing~\citep{muller2019does}, focal loss~\citep{lin2017focal}, and mixup~\citep{zhangMixupEmpiricalRisk2018}, which implicitly encourage calibrated predictions by modifying the loss. More recently, spectral-normalised neural Gaussian processes~\citep{liuSimplePrincipledUncertainty2020} have been proposed to enhance distance awareness in \gls*{dl}, potentially improving calibration. Ad hoc methods can yield better calibration but often require careful tuning and may increase training complexity.

\subsection{Platt and Temperature Scaling}

\begin{figure}[t]
  \centering
  \begin{subfigure}[t]{0.45\textwidth}
    \centering
    \includegraphics[width=\textwidth]{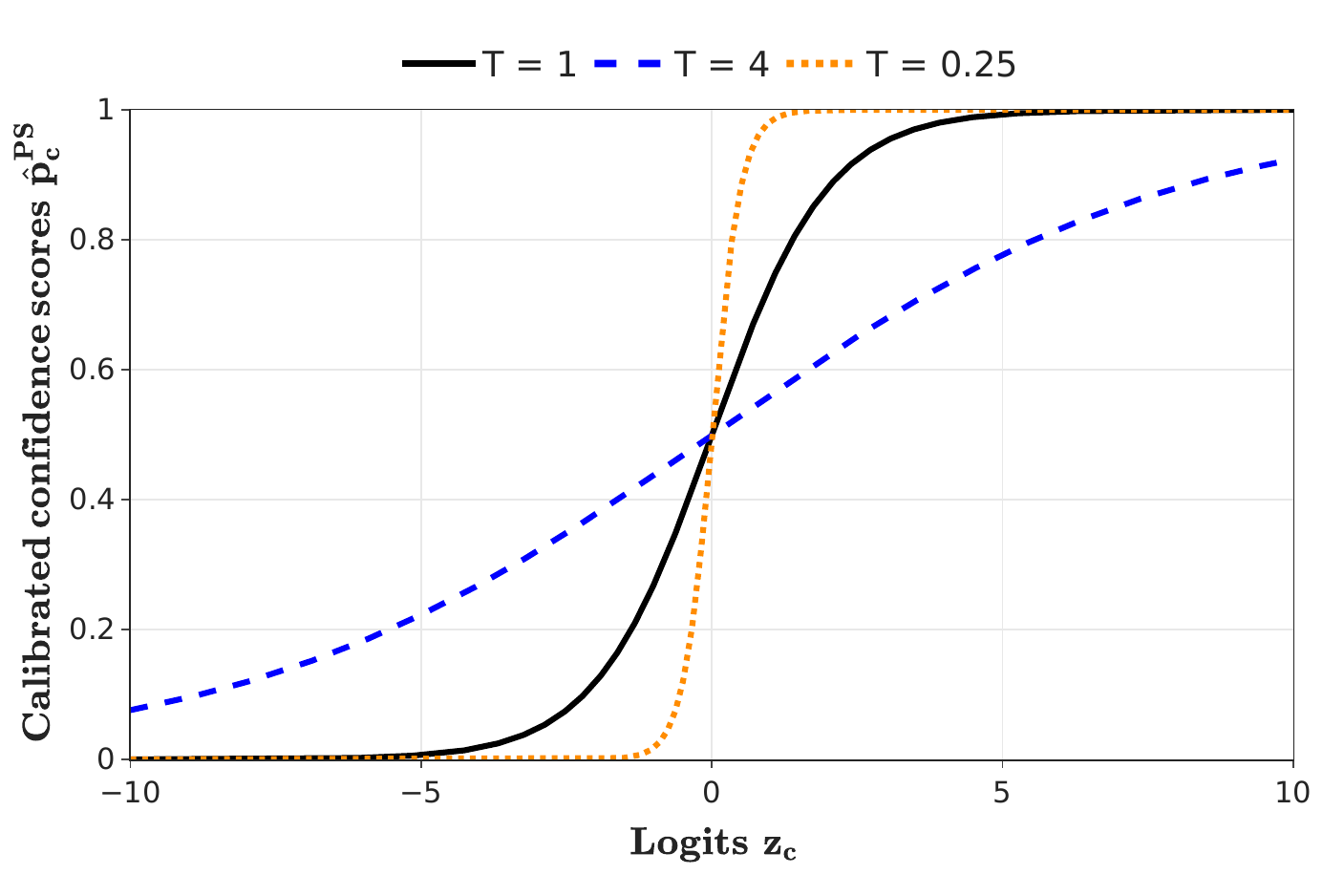}
    \subcaption{}\label{fig:platt_temp}
  \end{subfigure}\hfill
  \begin{subfigure}[t]{0.45\textwidth}
    \centering
    \includegraphics[width=\textwidth]{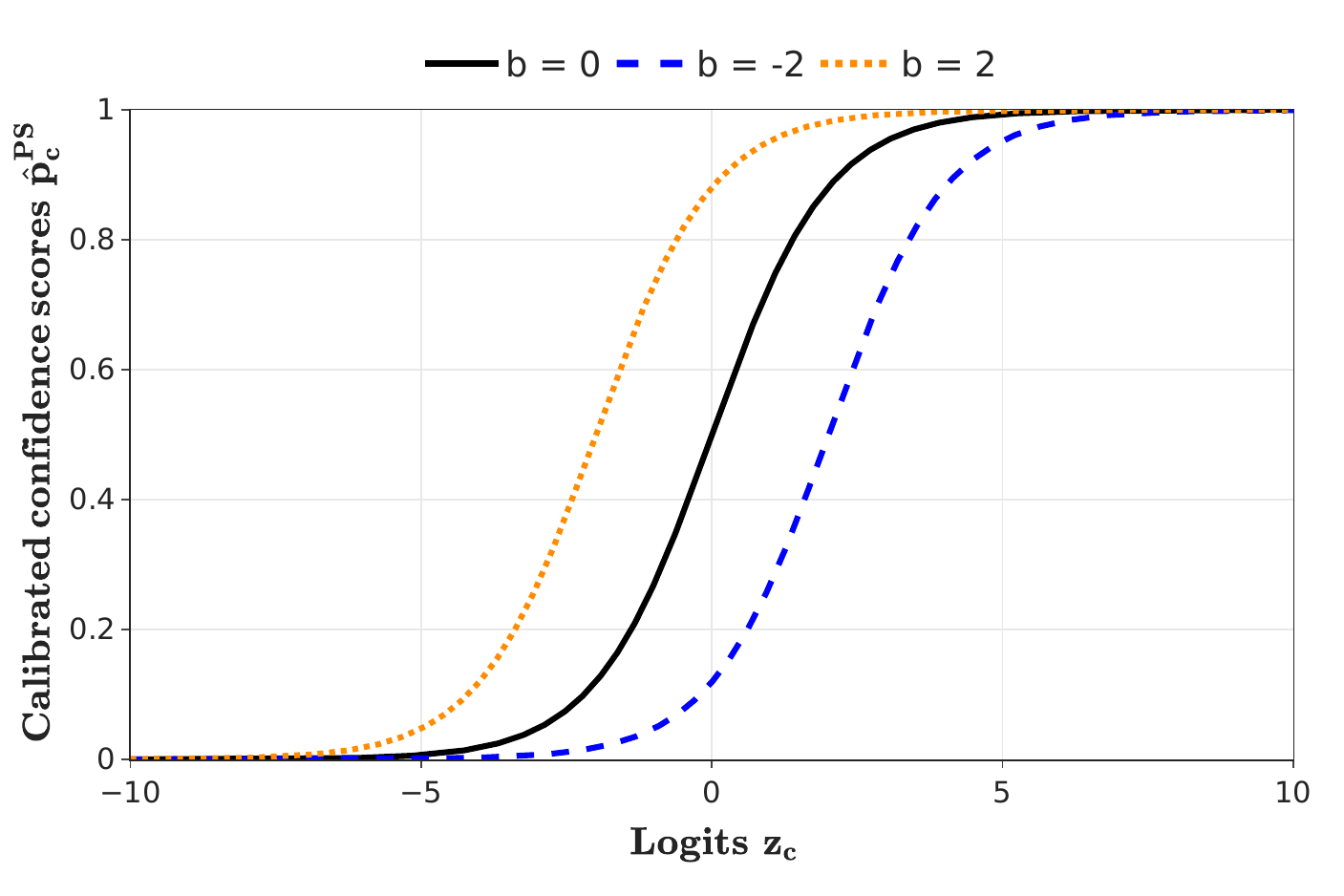}
    \subcaption{}\label{fig:platt_bias}
  \end{subfigure}\hfill
  \begin{subfigure}[t]{0.45\textwidth}
    \centering
    \includegraphics[width=\textwidth]{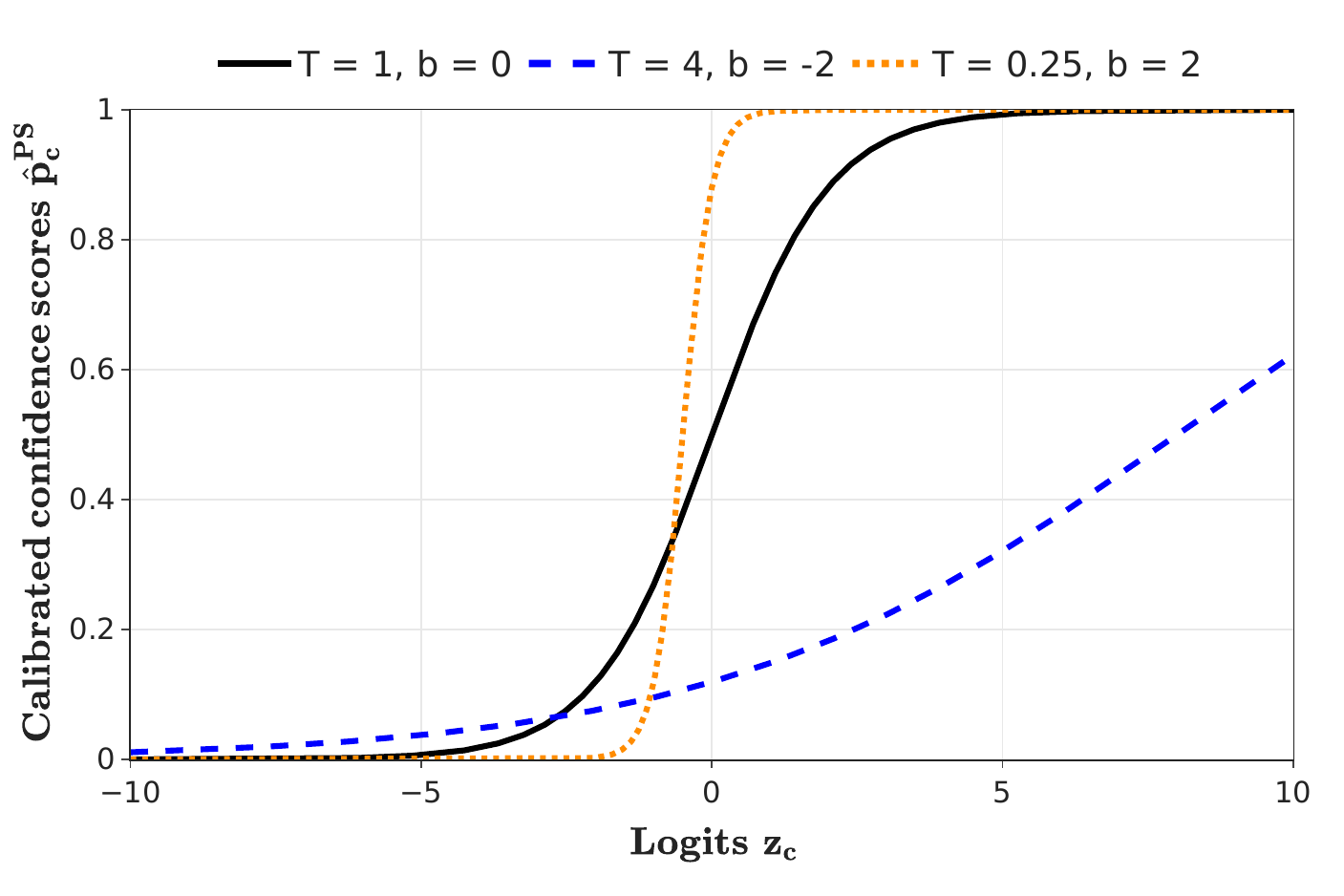}
    \subcaption{}\label{fig:platt_joint}
  \end{subfigure}
  \caption{Effects of Platt scaling on calibrated confidence scores $\hat{p}_c^{PS}$ as a function of logits $z_c$: (a) varying temperature parameter $T$ with fixed bias $b=0$, (b) varying bias $b$ with fixed temperature parameter $T=1$, and (c) joint variation of $T$ and $b$. The uncalibrated baseline $(T=1,\,b=0)$ is shown in black (solid) in all panels.}
  \label{fig:platt_scaling_overview}
\end{figure}

Modern bioacoustic models typically formulate animal sound detection as a multi-label classification problem~\citep{rauchBirdSetLargeScaleDataset2025,merrienboerPerch20Bittern2025,HEINRICH2025103081}.
For each class $c \in \{1,\dots,C\}$, the model produces a logit $z_c \in \mathbb{R}$, which is mapped to a confidence score $\hat p_c \in [0,1]$ via the sigmoid function, i.e., $\hat p_c = \sigma(z_c)$.
The resulting confidence scores can be calibrated post hoc without retraining the base model~\citep{platt1999probabilistic,niculescu2005predicting,guo2017calibration}.
A widely used method is \emph{\gls{ts}}~\citep{guo2017calibration}, which rescales the logits by a positive temperature parameter $T \in \mathbb{R}_{>0}$ and then re-applies the sigmoid:
\begin{equation}
  \hat p^{\mathrm{TS}}_c \;=\; \sigma\!\bigl(z_c/T\bigr), \qquad T>0.
\end{equation}
The parameter $T$ controls the sharpness of the resulting distribution of calibrated confidence scores, see Figure~\ref{fig:platt_temp}.
Setting $T=1$ recovers the uncalibrated model.
For $T>1$, confidences are smoothed.
Values below $0.5$ move up toward $0.5$, and values above $0.5$ move down toward $0.5$.
For $T<1$, the distribution is sharpened, pushing values below $0.5$ toward $0$ and values above $0.5$ toward $1$.
If the same temperature parameter is used for all classes $c$, \gls{ts} preserves the monotonic ordering of the model’s predicted confidence scores~\citep{guo2017calibration,tomani2022parameterized}.
Because $\sigma(0)=0.5$, confidence scores exactly at $0.5$ remain unchanged under \gls{ts}.
As a consequence, purely systematic offset errors (e.g., when the reliability curve~\citep{murphy1977reliability} lies consistently above or below the diagonal) cannot be corrected in this way~\citep{alexandari2020maximum}.
If offset errors need to be corrected, \emph{\gls{ps}}~\citep{platt1999probabilistic} provides a useful generalisation of \gls{ts}:
\begin{equation}
  \hat p^{\mathrm{PS}}_c \;=\; \sigma\!\bigl(z_c/T + b\bigr), \qquad T>0,\; b\in\mathbb{R}.
\end{equation}
In addition to the scaling parameter $T$, a bias $b \in \mathbb{R}$ is learned that shifts the entire sigmoid curve up or down, thereby changing the point at which a confidence score of $0.5$ is achieved (see Figures~\ref{fig:platt_bias} and \ref{fig:platt_joint}).
Positive biases ($b>0$) increase confidence scores, while negative biases ($b<0$) decrease them.
Both \gls{ts} and Platt scaling can also be parameterised on a per-class basis, using class-specific parameters $T_c$ and $b_c$ to compensate for class-dependent underconfidence and overconfidence~\citep{alexandari2020maximum,frenkel2021network}.
In practice, the parameters are estimated by minimising the \gls{nll} of the true labels on a held-out validation set~\citep{lin2007note,guo2017calibration}.
The resulting optimisation problems are convex and involve only a few parameters to learn~\citep{lin2007note,guo2017calibration}, making them well suited even when only limited data are available for calibration~\citep{niculescu2005predicting}.
After fitting, the learned parameters are applied to all future predictions without modifying the base model.

\subsection{Related work}
\label{sec:related-work}

\subsubsection{Uncertainty calibration in multi-label classification}

Multi-label classification presents unique calibration challenges due to label imbalance and the presence of co-occurring classes.
Recent research has highlighted that standard deep neural networks, when trained for multi-label tasks, often produce poorly calibrated probability estimates, especially in the presence of severe class imbalance and large label spaces. \citet{chengCalibratedMultilabelDeep} demonstrate that popular asymmetric loss functions, while effective for improving accuracy in unbalanced datasets, do not guarantee well-calibrated outputs. They attribute this to the lack of the strictly proper property in these losses and propose a new Strictly Proper Asymmetric loss, along with a label pair regulariser, to improve calibration without sacrificing accuracy. Their experiments show significant reductions in calibration error across diverse multi-label datasets.

In the context of extreme multi-label classification (XMLC), where the number of possible labels can reach into the millions, \citet{ullahLabelsExtremesHow2024} systematically evaluate the calibration of a wide range of XMLC models. They find that naive application of standard metrics such as ECE can be misleading in long-tailed label distributions. To address this, they introduce calibration@k metrics (e.g., ECE@k), which focus on the top-k predicted labels, providing a more meaningful assessment of calibration in XMLC scenarios. Furthermore, they show that post hoc calibration using isotonic regression can substantially improve reliability without degrading predictive performance, making it a practical tool for practitioners. In bioacoustics, the use of pretrained models (e.g. Perch v2~\citep{merrienboerPerch20Bittern2025}, BirdNET~\citep{kahlBirdNETDeepLearning2021}) is increasingly common due to a lack of labelled in-domain data. At the same time, the number of classes for which these models compute predictions is increasing, with models such as Perch v2 having almost 15,000 classes available.

\subsubsection{Uncertainty Calibration in Audio Classification}

Uncertainty calibration for deep audio classifiers is an emerging research area, motivated by the need for reliable confidence estimates in practical applications such as multimedia retrieval, urban sound monitoring, and bioacoustic surveys. While deep neural networks have achieved state-of-the-art performance in audio classification tasks~\citep{chenBEATsAudioPreTraining2023}, model calibration is not well investigated yet.

\citet{yeUncertaintyCalibrationDeep2022} provide one of the first systematic investigations of calibration methods for deep audio classifiers. They benchmark the calibration methods Monte Carlo Dropout, deep ensembles, focal loss, and spectral-normalised neural Gaussian processes (SNGP) on environmental sound and music genre classification datasets. Their results indicate that uncalibrated models are frequently overconfident, and that SNGP in particular achieves strong calibration and computational efficiency. Notably, while focal loss can improve calibration, it may degrade out-of-distribution detection performance, highlighting trade-offs between calibration and other uncertainty-related objectives.

In the context of bioacoustics, calibration is further complicated by the scarcity of strongly labelled data and the prevalence of rare species~\citep{stowell_computational_2022}. \citet{rizos2024propagating} propose a Bayesian variational method which propagates uncertainty (prediction and variance) using a Bayesian neural network and is used for post hoc label smoothing.

To the best of our knowledge, our work is the first to benchmark calibration of state-of-the-art bioacoustic classifiers.

\begin{table}[ht]
  \centering
  \caption{Overview of the different datasets in BirdSet~\citep{rauchBirdSetLargeScaleDataset2025}. J refers to the class imbalance index.}
  \label{tab:birdset}
  \setlength{\tabcolsep}{0.25em}
  \begin{adjustbox}{max width=0.5\textwidth}
    \begin{tabular}{p{0.8cm}|rrrrrr}
  \toprule
  \rowcolor{gray!15} \cellcolor{white}         & \textbf{Set}                                                     & \textbf{$\vert$Annotations$\vert$} & \textbf{$\vert$Duration$\vert$} & $\boldsymbol{J}$ & \#$\boldsymbol{C}$ \\
  \midrule
  \multirow{1}{*}{{\textit{Train}}}            &
  \href{https://xeno-canto.org/}{\texttt{XCL}} & 712,433                                                          & ~7,200h                            & 0.85                            & 9,734                                 \\
  \midrule
  \multirow{1}{*}{\textit{Val}}                & \href{https://zenodo.org/records/4656848}{\texttt{POW}}          & 16,052                             & 6.3h                            & 0.66             & 48                 \\
  \midrule
  \multirow{7}{*}{\textit{Test}}
  & \href{https://zenodo.org/records/7079124}{\texttt{PER}}          & 14,798                             & 21h                             & 0.78             & 132                \\
  & \href{https://zenodo.org/records/7525349}{\texttt{NES}}          & 6,952                              & 34h                             & 0.76             & 89                 \\
  & \href{https://zenodo.org/records/7078499}{\texttt{UHH}}          & 59,583                             & 50.9h                           & 0.64             & 25                 \\
  & \href{https://zenodo.org/records/7525805}{\texttt{HSN}}          & 10,296                             & 16.7h                           & 0.54             & 21                 \\
  & \href{https://github.com/fbravosanchez/NIPS4Bplus}{\texttt{NBP}} & 5,493                              & 0.8h                            & 0.92             & 51                 \\
  & \href{https://zenodo.org/records/7018484}{\texttt{SSW}}          & 50,760                             & 285h                            & 0.77             & 81                 \\
  & \href{https://zenodo.org/records/7050014}{\texttt{SNE}}          & 20,147                             & 33h                             & 0.70             & 56                 \\
  \bottomrule
\end{tabular}

  \end{adjustbox}
\end{table}
\section{\uppercase{Experiments}}
\label{sec:experiments}

\begin{table*}[ht]
  \centering
  \caption{Overview of key characteristics of bird sound classification models. For each model, we indicate the year of release, the number of classes the model is trained to classify, training method—supervised learning (SL) or self‐supervised learning (SSL)—as well as the architecture, number of parameters, embedding size, input duration (in seconds), input type and sample rate (in kHz).}
  \label{tab:modeldetails}
  \setlength{\tabcolsep}{0.25em}
  \begin{adjustbox}{max width=\textwidth}
    \begin{tabular}{lp{3cm} l c c c c c c c c }
  \toprule
  \rule{0pt}{2em} %

  & \multicolumn{1}{c}{ \textbf{Model}}
  & \multicolumn{1}{c}{ \textbf{Year}}
  & \multicolumn{1}{c}{\textbf{Classes}}
  & \multicolumn{1}{c}{
    \begin{tabular}[c]{@{}c@{}} \textbf{Training} \\ \textbf{Method}
  \end{tabular}}
  & \multicolumn{1}{c}{\textbf{Architecture}}
  & \multicolumn{1}{c}{
    \begin{tabular}[c]{@{}c@{}} \textbf{Parameters} \\ \textbf{\scriptsize(M)}
  \end{tabular}}
  & \multicolumn{1}{c}{
    \begin{tabular}[c]{@{}c@{}} \textbf{Embedding} \\ \textbf{Size}
  \end{tabular}}
  & \multicolumn{1}{c}{
    \begin{tabular}[c]{@{}c@{}} \textbf{Input} \\ \textbf{Duration \scriptsize(s)}
  \end{tabular}}
  & \multicolumn{1}{c}{
    \begin{tabular}[c]{@{}c@{}} \textbf{Input} \\ \textbf{Type}
  \end{tabular}}
  & \multicolumn{1}{c}{
    \begin{tabular}[c]{@{}c@{}} \textbf{Sample} \\ \textbf{Rate \scriptsize(kHz)}
  \end{tabular}}
  \\
  \midrule
  & \textbf{AudioProtoPNet-20}                                                                                          & \scriptsize2025 & 9,736  & SL       & ConvNeXt + Prototype Head        & 287  & 1024  & 5 & Spectrogram & 32 \\
  & \textbf{BirdMAE-XCL}                                                                                             & \scriptsize2025 & 9,735  & SSL + SL      & ViT-B + Linear Head              & 92.9 & 768 & 5 & Spectrogram & 32 \\
  & \textbf{ConvNeXt$_{BS}$}                                                                                         & \scriptsize2025 & 9,736  & SL       & ConvNeXt + Linear Head           & 88  & 768  & 5 & Spectrogram & 32 \\
  & \textbf{Perch v2}                                                                                                & \scriptsize2025 & 14,795 & SSL + SL & EfficientNet-B3 + Linear Head & 12  & 1536 & 5 & Spectrogram & 32 \\
  \bottomrule
\end{tabular}%

  \end{adjustbox}
\end{table*}

\subsection{Experimental Setup}
\subsubsection{Datasets}

While many datasets in bioacoustics exist, BirdSet~\citep{rauchBirdSetLargeScaleDataset2025} is not only widely used in the field~\citep{merrienboerPerch20Bittern2025,miron2025mattersbioacousticencoding,HEINRICH2025103081,schwinger2025foundation}, but also particularly well-suited for our experiments due to its multi-label nature and practical relevance. BirdSet consists of seven test datasets sourced from \gls{pam} deployments, covering a range of habitats and recording conditions. These seven test datasets are fully annotated by experts, while the training data is sourced from the public Xeno-Canto database. Table~\ref{tab:birdset} gives an overview of the datasets.

\subsubsection{Models -- Bird Sound Classifiers}

\paragraph{ConvNeXt$_{BS}$}

ConvNeXt$_{BS}$ is a ConvNeXt~\citep{liuConvNet2020s2022a} encoder with a linear classification head. The model weights were initialised using an ImageNet pretrained checkpoint and further trained via supervised learning on the BirdSet XCL training data. Extensive augmentations are used to allow the model to generalise better for the task of bird sound classification in \gls{pam} scenarios, see~\citep{rauchBirdSetLargeScaleDataset2025} for more details.

\paragraph{AudioProtoPNet}
AudioProtoPNet~\citep{HEINRICH2025103081} is an interpretable deep learning model for multi-label bird sound classification, adapted from the Prototypical Part Network (ProtoPNet) architecture~\citep{chen2019looks}. It leverages a ConvNeXt backbone to extract embeddings from input spectrograms and learns prototypical patterns for each bird species in the embedding space. During inference, the model classifies recordings by comparing them to learned prototypes, providing both accurate predictions and interpretable explanations for its decisions. Like ConvNeXt$_{BS}$, it is trained via supervision on the BirdSet XCL dataset. We use the largest version with 20 prototypes per class.

\paragraph{BirdMAE}
BirdMAE~\citep{rauchCanMaskedAutoencoders2025} is a domain-specialised masked autoencoder (MAE) designed for bird sound classification. Unlike general-purpose audio MAEs pretrained on AudioSet, BirdMAE is pretrained on the BirdSet XCL dataset to better capture the fine-grained acoustic characteristics of avian bioacoustics. The model supports both fine-tuning and parameter-efficient prototypical probing, where frozen representations are leveraged for downstream tasks. For this paper, we fine-tuned the base checkpoint on the BirdSet XCL training data using a linear classification head.

\paragraph{Perch v2}

Perch v2~\citep{merrienboerPerch20Bittern2025} is a state-of-the-art model for bioacoustic sound classification.
It builds upon the original Perch architecture (EfficientNet-B3) by incorporating additional training data and advanced augmentation techniques, resulting in improved performance across various evaluation metrics.

Table~\ref{tab:modeldetails} provides an overview of key model properties.
We excluded BirdNET~\citep{kahlBirdNETDeepLearning2021} from our experiments as its training data overlaps with the BirdSet evaluation datasets, which would bias our calibration assessment.

\subsubsection{Metrics}

\paragraph{class mean Average Precision (cmAP).} cmAP is obtained by computing the average precision (AP) for each class $c$ independently and then averaging these values across all $C$ classes, yielding a macro-average~\citep{rauchBirdSetLargeScaleDataset2025}. It can be computed by:
\begin{equation}
  \text{cmAP} := \frac{1}{C} \sum_{c=1}^{C} \text{AP}(c).
\end{equation}
This metric evaluates the model’s ability to consistently rank positive instances above negative ones across all possible decision thresholds, providing a threshold-independent measure of overall performance. Nonetheless, it can become noisy for classes with few positive labels, which reduces its comparability across datasets~\citep{denton2022improving}.

In a multi-label setting, reliability diagrams can be computed for each class individually and aggregated for simpler visualisation.

\paragraph{Expected Calibration Error (ECE).} ECE quantifies the difference between accuracy and confidences. The predictions are partitioned into $M$ bins (we use $M=15$ in our experiments). The error can be computed as follows:
\begin{equation}
  \text{ECE} = \sum_{m=1}^{M} \frac{|B_m|}{N} | \text{acc}(B_m) - \text{conf}(B_m) |
\end{equation}

where \(B_m\) is the set of indices of samples in bin \(m\), \(N\) is the total number of samples, \(\text{acc}(B_m)\) is the accuracy of predictions in bin \(m\), and \(\text{conf}(B_m)\) is the average confidence of predictions in bin \(m\).

\paragraph{Miscalibration Score (MCS).}  As ECE only quantifies the calibration error and gives no information on whether the  model is over- or underconfident, the MCS~\citep{aoTwoSidesMiscalibration2023} computes the difference between confidence and accuracy in each bin. A perfectly calibrated model has an MCS of $0$ while under-confident models have a negative score and overconfident models a positive score. The ECE equation is adapted to:
\begin{equation}
  \text{MCS} = \sum_{m=1}^{M} \frac{|B_m|}{N} ( \text{conf}(B_m) - \text{acc}(B_m) )
\end{equation}

As the MCS computes an average of the difference between $\text{conf}(B_m)$ and $\text{acc}(B_m)$, over- and underconfident bins can cancel each other out. Consequently, the MCS loses information if the model is miscalibrated in multiple regions. Therefore, we additionally evaluate:
\paragraph{Over-confidence Score (OCS).}
\begin{equation}
  \text{OCS} = \sum_{m=1}^{M} \frac{|B_m|}{N} max(\text{conf}(B_m) - \text{acc}(B_m),0)
\end{equation}
and
\paragraph{Under-confidence Score (UCS).}

\begin{equation}
  \text{UCS} = \sum_{m=1}^{M} \frac{|B_m|}{N} | min(\text{conf}(B_m) - \text{acc}(B_m),0) |
\end{equation}

These metrics quantify the extent of over- and underconfidence separately, providing a more nuanced view of calibration performance.

\paragraph{Multi-label adaption.} The calibration metrics ECE, MCS, OCS and UCS can be naively adapted for multi-label classification problems by treating each class as a binary classification problem and averaging the computed scores. As the samples are not uniformly distributed over each class, we weight them by the number of positive target labels.

\subsection{Benchmarking the Calibration of Bird Sound Classifiers}
In this section, we benchmark the calibration performance of multiple models using seven strongly labelled BirdSet test datasets and the POW validation set. We examine calibration at different levels of granularity, from global and per-dataset assessments to per-class and class-frequency levels, to determine how calibration varies across domains and classes.

\subsubsection{Experiment: Global and Per-dataset calibration}
\label{exp:global-and-per-dataset}

All test samples are collected to report the global performance for each model, neglecting the origin of the samples. Additionally, domain-specific calibration performance is investigated across individual BirdSet evaluation datasets. This experiment investigates calibration variability between domains (datasets) and how that compares to calibration metrics when aggregating across all domains.

\subsubsection{Experiment: Per-class calibration}
\label{exp:per-class}

\paragraph{Per-class Calibration.} We evaluate the per-class ECE given the number of samples per class, investigating calibration characteristics for each model on a class basis. This experiment investigates the generalisability of model calibration across imbalanced classes by evaluating how calibration metrics change when a class has more samples.

\paragraph{Calibration for Frequent and Rare Classes.} Additionally, we split the classes (species) of the BirdSet evaluation datasets into frequent and rare subsets. The frequent subset includes the 19 classes with the most samples per class, representing 50\% of the total samples. The rare subset includes samples from all other classes. Model calibration metrics are aggregated across these subsets to compare model performance differences for frequent and rare species.

\subsection{Post hoc Methods to Improve Calibration}

We also investigate whether calibration can be improved using the compute-efficient post hoc methods \gls{ts}~\citep{guo2017calibration} and \gls{ps}~\citep{platt1999probabilistic}. Both methods need specific data to optimise the scaling parameters. We experiment in two settings: (i) utilise the BirdSet POW dataset as a calibration set and (ii) leverage the first 10 minutes of each dataset.

As the classes of the POW dataset differ from the test datasets, only global scaling parameters (temperature, or temperature and bias) can be fitted, whereas using a part of the test dataset enables fitting per-class parameters. The test dataset is slightly reduced in this case, making a direct comparison between the two approaches difficult. Still, in practice it is often possible to label a small part of the test dataset. In both settings and for each method, we optimise the parameters using Adam~\citep{kingma2014adam} with a learning rate of $0.001$ for 1000 steps.

\section{\uppercase{Results and Discussion}}
\label{sec:results}
\begin{table}
  \centering
  \caption{The models AudioProtoPNet, BirdMAE, ConvNeXt$_{BS}$ and Perch v2 are evaluated on the seven BirdSet test sets and the POW validation set with the metrics cmAP, ECE, MCS, OCS and UCS. The "All" column reflects the class-wise average independent of dataset. Strongest performance is highlighted in \textbf{bold}. Arrows next to the metrics indicate what qualifies as good performance.}
  \label{tab:calibration-results}
  \setlength{\tabcolsep}{0.25em}
  \begin{adjustbox}{max width=0.48\textwidth}
    \renewcommand{\arraystretch}{0.55} %
\setlength{\tabcolsep}{2pt}

\begin{tabular}{p{2.5cm} | cccccccc !{\vrule width 1.3pt} c}
  \multicolumn{1}{c}{}                                                                        & \multicolumn{8}{c}{\textbf{BirdSet}}     &                                                                                                                                                                                                                                                                                                                                              \\
  \addlinespace[2pt]
  \cline{2-9}
  \addlinespace[2pt]
  \multicolumn{1}{c}{}                                                                        & \cellcolor{gray!25}\textbf{\textsc{PER}} & \cellcolor{gray!25}\textbf{\textsc{POW}} & \cellcolor{gray!25}\textbf{\textsc{NES}} & \cellcolor{gray!25}\textbf{\textsc{UHH}} & \cellcolor{gray!25}\textbf{\textsc{HSN}} & \cellcolor{gray!25}\textbf{\textsc{NBP}} & \cellcolor{gray!25}\textbf{\textsc{SSW}} & \cellcolor{gray!25}\textbf{\textsc{SNE}} & \cellcolor{gray!25}\textbf{All} \\
  \addlinespace[2pt]
  \cline{2-9}
  \addlinespace[2pt]
  \midrule
  \multicolumn{2}{l}{\vspace{0.25em}\hspace{0.25em}\textit{cmAP} \scriptsize(\(\uparrow\))}   &                                                                                                                                                                                                                                                                                                                                                                                         \\
  {\textbf{AudioProtoPNet}}                                                                   & \textbf{31.74}                           & 50.66                                    & 38.07                                    & 31.72                                    & \textbf{55.53}                           & \textbf{65.74}                           & 43.53                                    & \textbf{33.21}                           & \textbf{41.36}                  \\ [0.1em]
  {\textbf{BirdMAE}}                                                                          & 25.28                                    & 39.51                                    & 35.29                                    & 22.33                                    & 46.28                                    & 62.42                                    & 35.35                                    & 30.56                                    & 35.57                           \\ [0.1em]
  {\textbf{ConvNeXt$_{BS}$} }                                                                 & 17.60                                    & 33.99                                    & 34.14                                    & 24.72                                    & 48.72                                    & 61.36                                    & 34.79                                    & 30.05                                    & 32.20                           \\ [0.1em]
  {\textbf{Perch v2}}                                                                         & 22.74                                    & \textbf{51.74}                           & \textbf{39.29}                           & \textbf{36.71}                           & 50.75                                    & 65.64                                    & \textbf{45.25}                           & 33.15                                    & 39.16                           \\ [0.1em]
  \midrule
  \multicolumn{2}{l}{\vspace{0.25em}\hspace{0.25em}\textit{ECE} \scriptsize(\(\downarrow\))}  &                                                                                                                                                                                                                                                                                                                                                                                         \\
  {\textbf{AudioProtoPNet}}                                                                   & 15.12                                    & 24.15                                    & 13.00                                    & \textbf{14.23}                           & \textbf{10.29}                           & 16.16                                    & 13.12                                    & 14.34                                    & 5.25                            \\ [0.1em]
  {\textbf{BirdMAE}}                                                                          & 18.11                                    & \textbf{22.08}                           & 16.19                                    & 15.10                                    & 15.21                                    & 19.32                                    & 19.59                                    & 20.27                                    & 7.62                            \\ [0.1em]
  {\textbf{ConvNeXt$_{BS}$}  }                                                                & 11.13                                    & 24.51                                    & \textbf{2.46}                            & 16.13                                    & 11.81                                    & 3.28                                     & 1.90                                     & 6.99                                     & 1.10                            \\ [0.1em]
  {\textbf{Perch v2}}                                                                         & \textbf{10.98}                           & 22.46                                    & 2.54                                     & 15.93                                    & 10.89                                    & \textbf{3.04}                            & \textbf{1.37}                            & \textbf{6.04}                            & \textbf{0.92}                   \\ [0.1em]
  \midrule
  \multicolumn{2}{l}{\vspace{0.25em}\hspace{0.25em}\textit{MCS} \scriptsize(\(\rightarrow\))} &                                                                                                                                                                                                                                                                                                                                                                                         \\
  {\textbf{AudioProtoPNet} }                                                                  & \textbf{6.16}                            & -6.92                                    & 11.74                                    & \textbf{-1.19}                           & \textbf{1.14}                            & 14.19                                    & 11.93                                    & 7.96                                     & 4.43                            \\ [0.1em]
  {\textbf{BirdMAE}}                                                                          & 12.11                                    & \textbf{-2.71}                           & 15.48                                    & 1.83                                     & 8.96                                     & 17.86                                    & 18.78                                    & 15.62                                    & 7.01                            \\ [0.1em]
  {\textbf{ConvNeXt$_{BS}$}  }                                                                & -10.99                                   & -24.45                                   & \textbf{-2.38}                           & -16.07                                   & -11.80                                   & -3.23                                    & -1.86                                    & -6.99                                    & -1.09                           \\ [0.1em]
  {\textbf{Perch v2}}                                                                         & -10.93                                   & -22.46                                   & -2.53                                    & -15.93                                   & -10.88                                   & \textbf{-3.00}                           & \textbf{-1.35}                           & \textbf{-6.04}                           & \textbf{-0.91}                  \\ [0.1em]
  \midrule
  \multicolumn{2}{l}{\vspace{0.25em}\hspace{0.25em}\textit{OCS} \scriptsize(\(\downarrow\))}  &                                                                                                                                                                                                                                                                                                                                                                                         \\
  {\textbf{AudioProtoPNet}}                                                                   & 10.64                                    & 8.62                                     & 12.37                                    & 6.52                                     & 5.71                                     & 15.17                                    & 12.53                                    & 11.15                                    & 4.84                            \\ [0.1em]
  {\textbf{BirdMAE}}                                                                          & 15.11                                    & 9.68                                     & 15.84                                    & 8.47                                     & 12.09                                    & 18.59                                    & 19.18                                    & 17.95                                    & 7.32                            \\ [0.1em]
  {\textbf{ConvNeXt$_{BS}$} }                                                                 & 0.07                                     & 0.03                                     & 0.04                                     & 0.03                                     & \textbf{0.00}                            & \textbf{0.02}                            & 0.02                                     & \textbf{0.00}                            & 0.01                            \\ [0.1em]
  {\textbf{Perch v2}}                                                                         & \textbf{0.02}                            & \textbf{0.00}                            & \textbf{0.01}                            & \textbf{0.00}                            & 0.01                                     & \textbf{0.02}                            & \textbf{0.01}                            & \textbf{0.00}                            & \textbf{0.00}                   \\ [0.1em]
  \midrule
  \multicolumn{2}{l}{\vspace{0.25em}\hspace{0.25em}\textit{UCS} \scriptsize(\(\downarrow\))}  &                                                                                                                                                                                                                                                                                                                                                                                         \\
  {\textbf{AudioProtoPNet}  }                                                                 & 4.48                                     & 15.54                                    & 0.63                                     & 7.71                                     & 4.57                                     & 0.99                                     & 0.60                                     & 3.19                                     & 0.41                            \\ [0.1em]
  {\textbf{BirdMAE}}                                                                          & \textbf{3.00}                            & \textbf{12.40}                           & \textbf{0.35}                            & \textbf{6.63}                            & \textbf{3.12}                            & \textbf{0.73}                            & \textbf{0.40}                            & \textbf{2.32}                            & \textbf{0.30}                   \\ [0.1em]
  {\textbf{ConvNeXt$_{BS}$} }                                                                 & 11.06                                    & 24.48                                    & 2.42                                     & 16.10                                    & 11.80                                    & 3.26                                     & 1.88                                     & 6.99                                     & 1.10                            \\ [0.1em]
  {\textbf{Perch v2}}                                                                         & 10.95                                    & 22.46                                    & 2.53                                     & 15.93                                    & 10.89                                    & 3.02                                     & 1.36                                     & 6.04                                     & 0.91                            \\ [0.1em]
  \bottomrule
\end{tabular}

  \end{adjustbox}
\end{table}

\subsection{Benchmarking the Calibration of Bird Sound Classifiers}

\subsubsection{Global and per-dataset calibration}
Table~\ref{tab:calibration-results} summarises the results of the described metrics cmAP, ECE, MCS, OCS and UCS for each of the seven BirdSet test sets and the POW validation set as well as the class-wise average performance irrespective of dataset origin. All metrics are weighted by class frequency.

\paragraph{Discriminative Metrics}
Across all test samples, AudioProtoPNet achieves the strongest discriminative performance by cmAP with 41.36, followed by Perch v2 with 39.16. BirdMAE and ConvNeXt$_{BS}$ follow with 35.57 and 32.20, respectively. Both top-performing models set the highest scores on a per-dataset basis as well. Our results slightly differ from those reported by \citet{merrienboerPerch20Bittern2025}, since we use the larger version of AudioProtoPNet-20 and our BirdMAE model is fine-tuned on XCL rather than individually for each test dataset. Our BirdMAE checkpoint further utilises a simple classification head on top of the mean-pooled embeddings rather than the patch embeddings, which degrades performance.

\paragraph{Global calibration}

Considering the class-wise average independent of the test datasets in Table~\ref{tab:calibration-results}, AudioProtoPNet, which reaches the strongest performance by cmAP, is less calibrated than Perch v2 or ConvNeXt$_{BS}$.
Perch v2 appears to be calibrated best overall with an MCS of -0.91, suggesting slight underconfidence.
ConvNeXt$_{BS}$ performs similarly with an MCS of -1.09, also indicating slight underconfidence.
In contrast, AudioProtoPNet and BirdMAE are overconfident with an MCS of 4.43 and 7.01, respectively.
The misalignment of best performance by cmAP and MCS reiterates the importance of uncertainty calibration.

\paragraph{Cross-dataset variability}

\begin{figure}
  \includegraphics[width=\columnwidth]{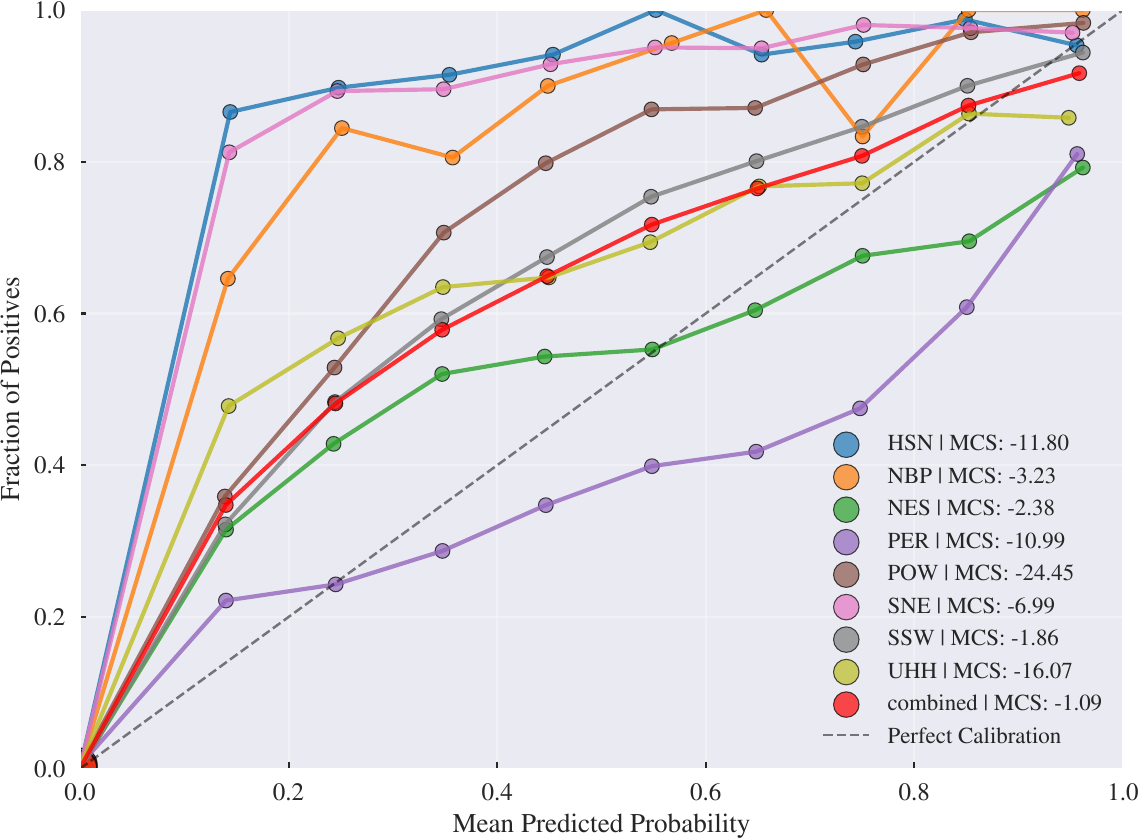}
  \caption{Demonstrating model (ConvNeXt$_{BS}$) calibration performance across individual BirdSet evaluation datasets and the combined performance when aggregated across all datasets.}
  \label{fig:reliability-ConvNeXt_BS}
\end{figure}

Considering ECE scores, Perch v2 and ConvNeXt$_{BS}$ achieve similar calibration scores with Perch v2 slightly outperforming ConvNeXt$_{BS}$ for all sets except for the NES test set.
For all test sets except for POW and UHH, BirdMAE achieves the highest ECE score, indicating poor calibration.
For all models, the combined ECE, aggregated across all BirdSet evaluation datasets, is better than the average of all datasets. Per-dataset ECE values for Perch v2 range from 1.37 at SSW to 22.46 at POW.
This demonstrates the importance of per-dataset evaluation, as calibration performance is potentially inflated when combining calibration metrics across distinct subdomains, masking the true calibration characteristics.

When analysing the miscalibration using MCS, Perch v2 achieves the best scores for NBP, SSW and SNE, while now AudioProtoPNet outperforms Perch v2 on PER.
Perch v2 and ConvNeXt$_{BS}$ are underconfident in all datasets whereas AudioProtoPNet and BirdMAE are overconfident except for POW and UHH (AudioProtoPNet) or POW (BirdMAE), respectively.

The consistent behaviour of underconfidence is in line with the reliability diagram in Figure~\ref{fig:reliability-diagram-combined}, in which Perch v2 and ConvNeXt$_{BS}$ exhibit values above the perfect calibration trajectory in all instances.
OCS scores further confirm that Perch v2 and ConvNeXt$_{BS}$ achieve nearly ideal performance with respect to overconfidence.
Subsequently, UCS scores carry the weight of the miscalibration for the two models.
While Figure~\ref{fig:reliability-diagram-combined} shows performance as a function of predicted probabilities and therefore does not have to align with the MCS scores, it does indicate whether the model exhibits a general over- or underconfident behaviour.
AudioProtoPNet and BirdMAE achieve consistently positive MCS scores.
Figure~\ref{fig:reliability-diagram-combined} shows that both models are overconfident for low predicted probabilities and underconfident for high predicted probabilities.

Figure~\ref{fig:reliability-ConvNeXt_BS} shows a reliability diagram for ConvNeXt$_{BS}$ over the seven test sets, the POW validation set, and the combined set.
Significant calibration variability across the evaluation datasets can be observed with most of the test set performances lying above the perfect calibration line, indicating underconfidence.
Figure~\ref{fig:reliability-ConvNeXt_BS} also highlights the potential misalignment of MCS values and reliability diagrams.
The reason for this is the very unequal distribution of samples in the predicted probability bins.
The UHH test set produces a reliability diagram which seems to be very similar to that of the SSW test set; however, the MCS values reach -16.07 and -1.86, respectively, suggesting a large difference in calibration.

Judging from the reliability diagram in Figure~\ref{fig:reliability-ConvNeXt_BS}, ConvNeXt$_{BS}$ is overconfident for the datasets PER and NES, which correspond to tropical locations, while the model is underconfident for other datasets.
While beyond the scope of this paper, a more detailed investigation into the number of samples per bin for specific datasets could disentangle why the reliability diagram suggests overconfidence and the MCS values suggest underconfidence for the PER and NES datasets.
In summary, a fine-grained evaluation of calibration metrics is a valuable tool to help understand model performance in different environments.

\begin{figure}
  \includegraphics[width=\columnwidth]{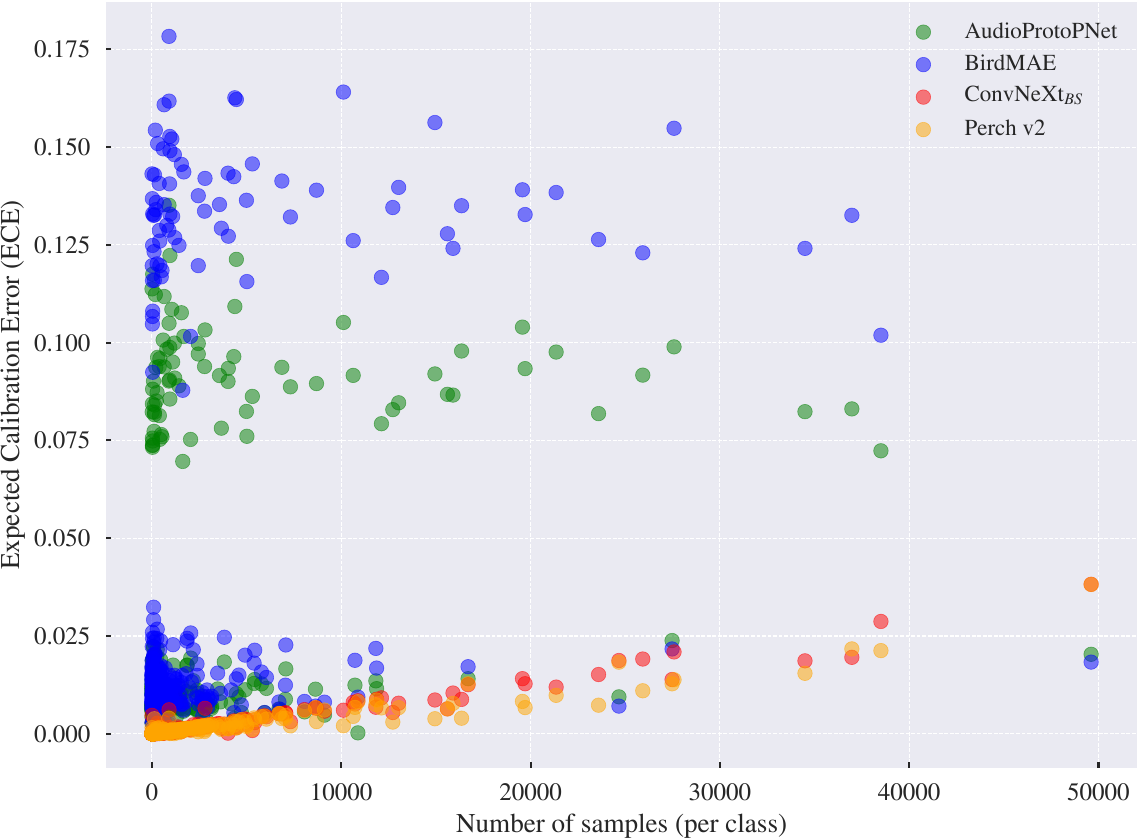}
  \caption{Class-wise ECE scores over number of samples per class for different models shown by colour.}
  \label{fig:ece-vs-samplecount}
\end{figure}

\subsubsection{Per-class calibration}

\paragraph{Per-class calibration}

Figure~\ref{fig:ece-vs-samplecount} shows the ECE scores per class over the number of samples in each class. Different models are shown by colour.
The well-performing models by ECE according to Table~\ref{tab:calibration-results} are Perch v2 and ConvNeXt$_{BS}$, which show very low ECE scores that exhibit a linear increase with the number of samples per class.
AudioProtoPNet and BirdMAE, on the other hand, show less consistent ECE values and both models seem to exhibit a distinct split between many classes reaching low ECE values ($ECE < 0.04$) and higher ECE values ($ECE > 0.06$).
For the largest class, AudioProtoPNet and BirdMAE reach ECE values below Perch v2 and ConvNeXt$_{BS}$.

The very large spread of ECE values for the underrepresented classes in AudioProtoPNet and BirdMAE provides an explanation for the poor performance per dataset in the previous section.
Well-calibrated classes can be masked by poorly calibrated classes, resulting in low overall calibration performance.
The distinct difference in class-wise performance could be influenced by some classes occurring only in very challenging noise conditions, resulting in an overall underperformance for that class.
It further highlights that calibration scores can vary greatly between classes of very similar sizes.

\paragraph{Common and Rare Species Calibration}

\begin{figure}
  \includegraphics[width=\columnwidth]{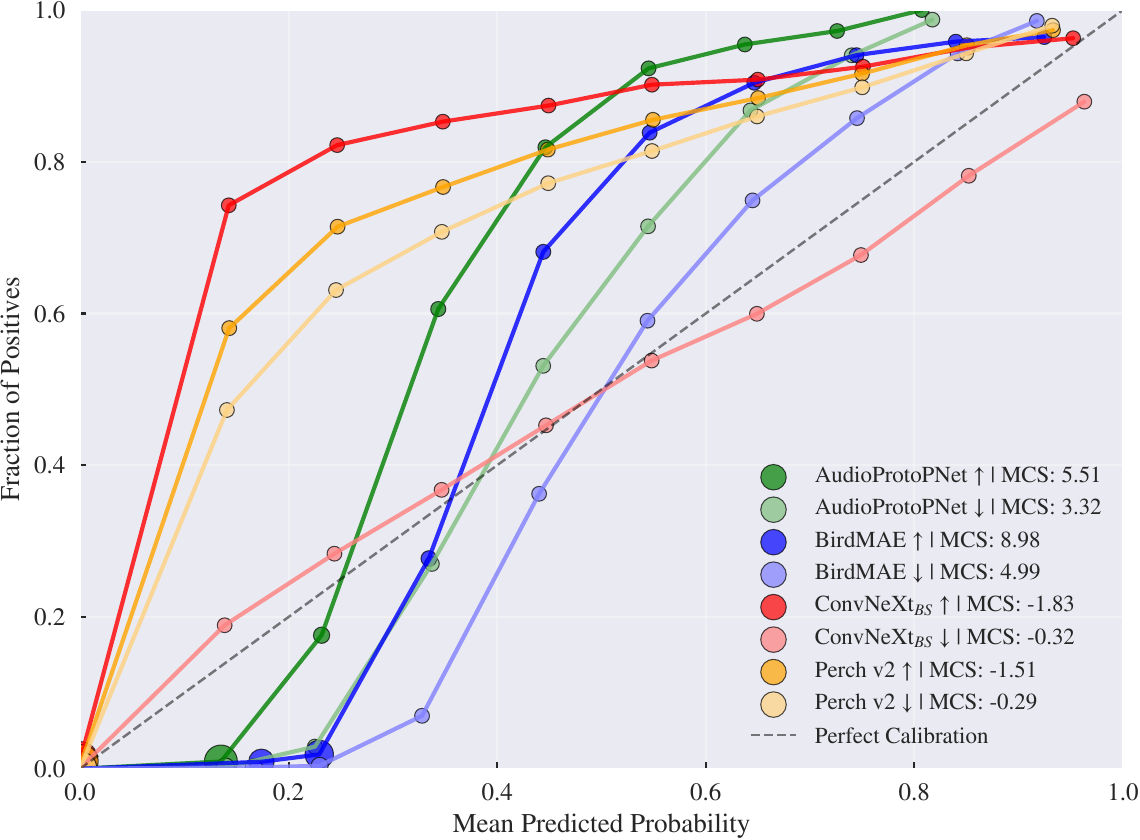}
  \caption{Reliability diagrams investigating calibration for common $\uparrow$ and rare $\downarrow$ subsets for different models. MCS value for all models and subsets are noted.}
  \label{fig:reliability-diagram-top-bottom}
\end{figure}

As visualised in Figure~\ref{fig:reliability-diagram-top-bottom}, Perch v2 and ConvNeXt$_{BS}$ trace above the perfect calibration line for both common and rare subsets, indicating underconfidence across most probability bins. Their OCS remains close to zero in both subsets, and miscalibration is dominated by the UCS. Interestingly, the rare subset shows a slight shift toward higher apparent overconfidence at very low confidence bins, but the overall MCS stays negative, consistent with the global pattern. This aligns with the class-wise view in Figure~\ref{fig:ece-vs-samplecount}, where these two models show low and relatively well-structured ECE that scales with sample count.

By contrast, AudioProtoPNet and BirdMAE display larger calibration variability between common and rare classes. For rare species, the reliability curves bend further away from the identity line at low probabilities (indicating stronger overconfidence), while at higher probabilities they get closer to the line (less underconfidence). This pattern yields positive MCS for the rare subset and increases OCS—consistent with their broader per-class ECE spread in Figure~\ref{fig:ece-vs-samplecount}. In practice, this suggests that rare classes are particularly susceptible to overconfident predictions in these two models, which can be problematic for downstream thresholding and decision support.

We note that we define rare species solely from the test sets; further investigation could take the train set distribution into consideration. In addition, the used metrics are sensitive to individual samples, when only a few instances per class are available.

\begin{table*}[t]
  \centering
  \caption{Post hoc calibration results using temperature scaling (TS) and Platt scaling (PS) on BirdSet test sets measured with MCS. 'TS POW' and 'PS POW' indicate global scaling parameters fitted on the POW dataset, while 'TS per class' and 'PS per class' indicate per-class scaling parameters fitted on the first 10 minutes of each test dataset. The 'Base' row reflects performance without post hoc calibration on the remaining test set. As a subscript the relative performance improvement is marked; for the 'Base' row the absolute difference to the full test set is given. Best results are highlighted in \textbf{bold}.}
  \label{tab:posthoc-results}
  \setlength{\tabcolsep}{0.25em}
  \begin{adjustbox}{max width=\textwidth}
    \renewcommand{\arraystretch}{0.55} %
\setlength{\tabcolsep}{2pt}

\begin{tabular}{p{2.5cm} | c | ccccccc }
  \multicolumn{1}{c}{}                                                       & \multicolumn{7}{c}{\textbf{BirdSet}}                                                                                                                                                                                                                                                                                                                          \\
  \addlinespace[2pt]
  \cline{2-9}
  \addlinespace[2pt]
  \multicolumn{1}{c}{}                                                       & \cellcolor{gray!25}\textbf{\textsc{POW$_{val}$}} & \cellcolor{gray!25}\textbf{\textsc{PER}} & \cellcolor{gray!25}\textbf{\textsc{NES}} & \cellcolor{gray!25}\textbf{\textsc{UHH}} & \cellcolor{gray!25}\textbf{\textsc{HSN}} & \cellcolor{gray!25}\textbf{\textsc{NBP}} & \cellcolor{gray!25}\textbf{\textsc{SSW}} & \cellcolor{gray!25}\textbf{\textsc{SNE}} \\
  \addlinespace[2pt]
  \cline{2-9}
  \addlinespace[2pt]
  \midrule
  \multicolumn{2}{l}{\vspace{0.25em}\hspace{0.25em}\textbf{AudioProtoPNet}}  &                                                                                                                                                                                                                                                                                                                                                               \\
  {\textit{TS POW}}                                                          & $-19.79_{-186\%}$                                & $-6.62_{-7.5\%}$                         & $\textbf{0.21}_{+98\%}$                  & $-12.35_{-938\%}$                        & $-8.50_{-646\%}$                         & $1.14_{+92\%}$                           & $0.57_{+95\%}$                           & $-3.53_{+56\%}$                          \\ [0.1em]
  {\textit{PS POW}}                                                          & $-17.61_{-154\%}$                                & $-7.07_{-15\%}$                          & $-0.71_{+94\%}$                          & $-12.41_{-943\%}$                        & $-8.36_{-633\%}$                         & $\textbf{0.43}_{+97\%}$                  & $\textbf{-0.34}_{+97\%}$                 & $-3.49_{+56\%}$                          \\ [0.1em]
  \cmidrule(lr){1-9}
  {\textit{Base} }                                                           & $-6.91_{+0.01}$                                  & $6.16_{+0.00}$                           & $11.74_{+0.00}$                          & $-1.20_{-0.01}$                          & $1.13_{-0.01}$                           & $14.05_{-0.14}$                          & $11.93_{+0.00}$                          & $7.95_{-0.01}$                           \\ [0.1em]
  {\textit{TS per class}}                                                    & $-9.57_{-38\%}$                                  & $-1.56_{+75\%}$                          & $-0.88_{+93\%}$                          & $-2.96_{-147\%}$                         & $-4.39_{-288\%}$                         & $-1.20_{+91\%}$                          & $-1.36_{+89\%}$                          & $-2.82_{+65\%}$                          \\ [0.1em]
  {\textit{PS per class}}                                                    & $\textbf{-0.36}_{+95\%}$                         & $\textbf{-0.03}_{+100\%}$                & $-0.52_{+96\%}$                          & $\textbf{-0.72}_{+40\%}$                 & $\textbf{-0.46}_{+59\%}$                 & $-0.55_{+96\%}$                          & $-1.19_{+90\%}$                          & $\textbf{-0.63}_{+92\%}$                 \\ [0.1em]
  \midrule
  \multicolumn{2}{l}{\vspace{0.25em}\hspace{0.25em}\textbf{BirdMAE}}         &                                                                                                                                                                                                                                                                                                                                                               \\
  {\textit{TS POW}}                                                          & $-16.01_{-491\%}$                                & $-3.00_{+75\%}$                          & $1.70_{+89\%}$                           & $-11.77_{-543\%}$                        & $-5.36_{+40\%}$                          & $2.73_{+85\%}$                          & $3.56_{+81\%}$                           & $\textbf{0.62}_{+96\%}$                  \\ [0.1em]
  {\textit{PS POW}}                                                          & $-17.91_{-561\%}$                                & $-5.48_{+55\%}$                          & $\textbf{-0.06}_{+100\%}$                & $-13.52_{-639\%}$                        & $-7.41_{+17\%}$                          & $0.37_{+98\%}$                           & $1.22_{+94\%}$                           & $-1.78_{+89\%}$                          \\ [0.1em]
  \cmidrule(lr){1-9}
  {\textit{Base}}                                                            & $-2.70_{+0.01}$                                  & $12.11_{+0.00}$                          & $15.48_{+0.00}$                          & $1.83_{+0.00}$                           & $8.95_{-0.01}$                           & $17.82_{-0.04}$                          & $18.78_{+0.00}$                          & $15.62_{+0.00}$                          \\ [0.1em]
  {\textit{TS per class}}                                                    & $-7.93_{-194\%}$                                 & $-0.76_{+94\%}$                          & $-0.49_{+97\%}$                          & $-2.46_{-34\%}$                          & $-3.41_{+62\%}$                          & $-0.70_{+96\%}$                          & $\textbf{-1.16}_{+94\%}$                 & $-2.15_{+86\%}$                          \\ [0.1em]
  {\textit{PS per class}}                                                    & $\textbf{0.39}_{+86\%}$                          & $\textbf{0.16}_{+99\%}$                  & $-0.58_{+96\%}$                          & $\textbf{-0.34}_{+81\%}$                 & $\textbf{-1.15}_{+87\%}$                 & $\textbf{-0.32}_{+98\%}$                 & $-1.22_{+94\%}$                          & $-0.75_{+95\%}$                          \\ [0.1em]
  \midrule
  \multicolumn{2}{l}{\vspace{0.25em}\hspace{0.25em}\textbf{ConvNeXt$_{BS}$}} &                                                                                                                                                                                                                                                                                                                                                               \\
  {\textit{TS POW }}                                                         & $-22.31_{+8.8\%}$                                & $-9.94_{+9.6\%}$                         & $-1.75_{+26\%}$                          & $-15.55_{+3.2\%}$                        & $-11.30_{+4.2\%}$                        & $-2.10_{+35\%}$                          & $-1.13_{+39\%}$                          & $-6.39_{+8.6\%}$                         \\ [0.1em]
  {\textit{PS POW  }}                                                        & $-18.59_{+24\%}$                                 & $-9.05_{+18\%}$                         & $-1.39_{+42\%}$                          & $-14.75_{+8.2\%}$                        & $-10.66_{+9.7\%}$                        & $-1.34_{+59\%}$                          & $\textbf{-0.72}_{+61\%}$                 & $-5.57_{+20\%}$                          \\ [0.1em]
  \cmidrule(lr){1-9}
  {\textit{Base  }}                                                          & $-24.46_{-0.01}$                                 & $-10.99_{+0.00}$                         & $-2.39_{-0.01}$                          & $-16.08_{-0.01}$                         & $-11.82_{-0.02}$                         & $-3.29_{-0.06}$                          & $-1.86_{+0.00}$                          & $-6.99_{+0.00}$                          \\ [0.1em]
  {\textit{TS per class } }                                                  & $\textbf{-9.72}_{+60\%}$                         & $\textbf{-1.45}_{+87\%}$                 & $\textbf{-0.79}_{+67\%}$                 & $\textbf{-4.10}_{+75\%}$                 & $\textbf{-3.71}_{+69\%}$                 & $-1.27_{+61\%}$                          & $-1.39_{+25\%}$                          & $\textbf{-2.39}_{+66\%}$                 \\ [0.1em]
  {\textit{PS per class }  }                                                 & $-18.76_{+23\%}$                                 & $-9.55_{+13\%}$                          & $-2.11_{+12\%}$                          & $-15.45_{+3.9\%}$                        & $-11.51_{+2.6\%}$                        & $\textbf{-1.03}_{+69\%}$                 & $-1.83_{+1.6\%}$                         & $-4.58_{+34\%}$                          \\ [0.1em]
  \midrule
  \multicolumn{2}{l}{\vspace{0.25em}\hspace{0.25em}\textbf{Perch v2}}        &                                                                                                                                                                                                                                                                                                                                                               \\
  {\textit{TS POW} }                                                         & $-24.39_{-8.6\%}$                                & $-10.43_{+4.6\%}$                        & $-1.48_{+42\%}$                          & $-10.93_{+31\%}$                         & $-6.33_{+42\%}$                          & $-2.01_{+33\%}$                          & $-0.63_{+53\%}$                          & $-5.58_{+7.6\%}$                          \\ [0.1em]
  {\textit{PS POW}}                                                          & $-26.13_{-16\%}$                                 & $-10.69_{+2.2\%}$                        & $-1.70_{+33\%}$                          & $-12.92_{+19\%}$                         & $-8.29_{+24\%}$                          & $-2.19_{+27\%}$                          & $-0.89_{+34\%}$                          & $-6.45_{-6.8\%}$                         \\ [0.4em]
  \cmidrule(lr){1-9}
  {\textit{Base}}                                                            & $-22.46_{+0.00}$                                 & $-10.93_{+0.00}$                         & $-2.53_{+0.00}$                          & $-15.93_{+0.00}$                         & $-10.89_{-0.01}$                         & $-3.02_{-0.02}$                          & $-1.35_{+0.00}$                          & $-6.04_{+0.00}$                          \\ [0.1em]
  {\textit{TS per class}}                                                    & $-26.71_{-19\%}$                                 & $-10.89_{+0.4\%}$                        & $-1.72_{+32\%}$                          & $-12.33_{+23\%}$                         & $-7.78_{+29\%}$                          & $-2.15_{+29\%}$                          & $-1.08_{+20\%}$                          & $-6.90_{-14\%}$                          \\ [0.1em]
  {\textit{PS per class}}                                                    & $\textbf{-14.05}_{+37\%}$                        & $\textbf{-6.96}_{+36\%}$                 & $\textbf{-0.36}_{+86\%}$                 & $\textbf{-2.24}_{+86\%}$                 & $\textbf{4.24}_{+61\%}$                  & $\textbf{-1.88}_{+38\%}$                 & $\textbf{0.07}_{+95\%}$                  & $\textbf{-2.47}_{+59\%}$                 \\ [0.1em]
  \bottomrule
\end{tabular}

  \end{adjustbox}
\end{table*}

\subsection{Temperature and Platt Scaling}

Table~\ref{tab:posthoc-results} summarises the MCS results for \gls{ts} and \gls{ps} using two different calibration set settings: \textit{Global} parameters fitted on POW and \textit{Per-class} fitted on the first 10 min of each test dataset.

\paragraph{Global}
Results for global \gls{ts} are mixed. While ConvNeXt$_{BS}$ and Perch v2 show improvements in MCS on all test datasets ranging from 3.2\% to 53\%, AudioProtoPNet and BirdMAE exhibit more varied performance, with some datasets showing degradation in MCS by over 938\%. On some datasets the improvements are substantial, with up to 98\%. For AudioProtoPNet, \gls{ts} seems to degrade calibration substantially on UHH and HSN and on PER (where the small relative change masks a sign flip from overconfident to underconfident). BirdMAE improves except for UHH. \Gls{ps} follows the same trend as \gls{ts}, with slightly better performance in most cases.

\paragraph{Per-class}

The difference of MCS on the remaining test set (- first 10 min) compared to the full test set is marginal for all models. Comparing global and per-class scaling parameters is therefore valid. Per-class \gls{ts} shows improvements for all models and datasets except for AudioProtoPNet on UHH, POW, and HSN, BirdMAE on POW and UHH and Perch v2 on POW and SNE. Here, also substantial degradations of up to -288\% are observed. Improvements range up to 97\%. Per-class \gls{ps} improves MCS for nearly all model--dataset combinations, with improvements up to 100\%, and typically outperforms \gls{ts}.

Both post hoc calibration methods demonstrate that substantial improvements to the deployment-specific calibration can be achieved with minor computational requirements. For example, AudioProtoPNet appears better calibrated than Perch v2 after per-class \gls{ps}.
However, no model or post hoc method consistently outperforms the others across all datasets, which reiterates that calibration and improvements using post hoc methods are domain-specific.

\subsection{Implications for deployment}
From a practitioner’s perspective, these observations have several implications:
\begin{enumerate}
  \item Thresholds tuned on common classes can be misaligned for rare classes, especially for models with higher OCS on rare species.
  \item Calibration evaluations between per-class and per-dataset are likely to differ which should be considered when deciding which calibration to prioritise.
  \item Lightweight post hoc calibration (Table~\ref{tab:posthoc-results}) can be prioritised for the rare subset, either via per-class scaling or via subgroup-aware global parameters, to rein in rare-class overconfidence without materially affecting the common subset.
  \item When communicating uncertainty to end users (e.g., survey practitioners), reporting OCS/UCS alongside ECE or MCS helps reveal whether observed errors stem from systematic over- or underconfidence and whether that profile differs between common vs. rare species.
\end{enumerate}

\section{\uppercase{Conclusions}}
\label{sec:conclusion}

We presented a first comprehensive benchmark of uncertainty calibration for four multi-label deep learning bird sound classifiers across seven expert-annotated BirdSet test datasets and the POW validation set. Beyond reporting class-wise mean average precision (cmAP), we evaluated calibration holistically using reliability diagrams and threshold-free metrics: expected calibration error (ECE), miscalibration score (MCS), and the decomposition into over and underconfidence (OCS/UCS). We analysed calibration at three granularities—global, per-dataset, and per-class—and examined behaviour for common versus rare species. Finally, we assessed lightweight post hoc methods (temperature and Platt scaling) fitting global calibration parameters on the POW validation set and per-class parameters on the first 10 min of each test dataset.

Three findings stand out. First, calibration varies substantially across datasets and classes; aggregating across domains can mask systematic miscalibration. In our setting, Perch v2 and ConvNeXt$_{BS}$ are consistently underconfident (near-zero OCS, UCS-dominated MCS), whereas AudioProtoPNet and BirdMAE show mixed behaviour with stronger overconfidence at low probabilities and underconfidence at higher probabilities. Second, calibration is also influenced by the frequency of classes. We demonstrate a shift towards model overconfidence when aggregated across less-frequent classes. Also surprisingly, class-wise calibration appears better for less-frequent classes. This may point to potential bias when estimating calibration with limited samples. Investigating the reliability and bias of calibration metrics for rare classes remains an important direction for future work. Third, simple post hoc calibration can help, but its efficacy is model- and subset-dependent: global scaling tended to improve Perch v2 and ConvNeXt$_{BS}$, while per-class Platt / temperature scaling often brought the largest gains, alongside occasional degradations that call for careful validation.

Our analysis focuses on post hoc calibration using limited in-domain data. Ad hoc training strategies and Bayesian approaches may further improve distance awareness and reliability. Although our metrics reduce threshold dependence, ECE and related scores can be unstable when there are few positives for rare species. Bootstrapped confidence intervals and calibration@k metrics for extreme multi-label settings are promising extensions. Finally, evaluating additional bioacoustic datasets and deployment scenarios (devices, habitats, noise regimes) will strengthen external validity and support robust thresholding policies for conservation monitoring.

\section*{\uppercase{Acknowledgements}}

This research has been funded by the German Ministry for the Environment, Nature Conservation, Nuclear Safety, and Consumer Protection through the project "DeepBirdDetect - Automatic Bird Detection of Endangered Species Using Deep Neural Networks" (67KI31040C).

\bibliographystyle{apalike}
{\small
\bibliography{references}}

\end{document}